\documentclass[]{article}

\usepackage{amsmath,amssymb,amsthm} % AMS packages
\usepackage{mathrsfs,mathtools} % additional math packages
\usepackage{algorithm,algorithmic} % for pseudo-code
\usepackage{authblk} % for author block

\newfloat{procedure}{tbp}{lop}
\floatname{procedure}{Procedure}

\usepackage{hyperref}
\hypersetup{colorlinks,urlcolor=blue}
\hypersetup{
    colorlinks,
    linkcolor={blue},
    citecolor={blue},
    urlcolor={blue}
}

\title{Initialization-robust characterization of deep sub-electron read noise pixels via annealed PCH-EM}
\author[1]{Aaron J.~Hendrickson}
\author[2]{David P. Haefner}
\affil[1]{U.S.~Navy, NAWCAD DAiTA Group, Patuxent River, MD 20670}
\affil[2]{U.S.~Army, DEVCOM C5ISR Center, Fort Belvoir, VA 22060}

\begin{document}

\maketitle

\begin{abstract}
We present an annealed Photon Counting Histogram Expectation Maximization (PCH-EM) algorithm for maximum likelihood characterization of Deep Sub-Electron Read Noise (DSERN) pixels. The annealed variant mitigates initialization-dependent convergence to suboptimal local optima of the likelihood function while achieving uncertainties substantially lower than the Photon Transfer (PT) method in the DSERN regime. Although the annealing principle is optimizer-agnostic, we pair it with PCH-EM for tractability; a single temperature parameter and simple cooling schedule suffice, without re-deriving the original EM update equations. Simulations across varied starting points show more stable parameter estimates with equal or better final likelihoods than the baseline. While designed for DSERN, the method applies across read-noise regimes and matches PT performance outside DSERN. Practically, the method enables reliable characterization of DSERN devices, including direct calibration from raw gray counts to electron counts for photon number resolving applications.
\end{abstract}

\begin{table}[b]\footnotesize\hrule\vspace{1mm}
Send correspondence to A.J.H: aaron.j.hendrickson2.civ@us.navy.mil
\end{table}

%%%%%%%%%%%%%%%%%%%%%%%%%%%%%%%%%%%%%%%%%%%%%%%%%%%%%%%%%%%%%%%%%%%%%%%%%%%%%%%%%%%%
%%%%%%%%%%%%%%%%%%%%%%%%%%%%%%%%%%%%%%%%%%%%%%%%%%%%%%%%%%%%%%%%%%%%%%%%%%%%%%%%%%%%
%%%%%%%%%%%%%%%%%%%%%%%%%%%%%%%%%%%%%%%%%%%%%%%%%%%%%%%%%%%%%%%%%%%%%%%%%%%%%%%%%%%%

\section{Introduction}
\label{sec:Introduction}

The Expectation-Maximization (EM) algorithm remains one of the most widely used approaches for Maximum Likelihood Estimation (MLE) in latent variable models \cite{dempster_1977}. Its generality and simplicity have made it a cornerstone in applications ranging from machine learning to image sensor characterization. The central practical limitation in non-convex settings is initialization-dependent convergence: EM monotonically converges to a stationary point of the likelihood function determined by its starting values, so different initializations can converge to different local optima, not the global maxima of interest in MLE. This phenomenon is not unique to EM but common to most local optimization methods on non-convex landscapes.

In prior work, we introduced the \emph{single-sample} Photon Counting Histogram EM (PCH-EM) algorithm as an implementation of EM specific to the characterization of image sensor pixels exhibiting Deep Sub-Electron Read Noise (DSERN)  \cite{hendrickson_2023_PCHEM_theory, hendrickson_2023_PCHEM_verification}. This single-sample implementation yielded estimates of key pixel parameters, e.g., DC offset, conversion gain, and read noise, with substantially lower uncertainty compared to the Photon Transfer (PT) method in the DSERN regime \cite{janesick_2007}. This lower estimation uncertainty is not only useful for characterization but also for calibrating the raw output of DSERN capable devices into quanta (electron) counts. Specifically, the quanta counting accuracy of DSERN pixels is theoretically limited by the magnitude of read noise however this theoretical limit is further deteriorated by errors in the parameter estimates since the estimates are utilized in the calibration process \cite{nakamoto_2022,hendrickson_2024_PCA}.

A practical limitation of the single-sample PCH-EM approach for CMOS Quanta Image Sensors (QIS) was that it applied only to pixels in the DSERN regime. Because per-pixel read noise in CMOS devices is often heavy-tailed \cite{fowler_2013}, a nontrivial fraction of pixels typically fall outside DSERN, making the single-sample method insufficient for full-array characterization. This was addressed in \cite{hendrickson_2024} with a \emph{multi-sample} extension of PCH-EM that incorporates experimental data across a range of signal (quanta exposure) levels. Spreading measurements over signal levels supplies complementary, signal-dependent information: data captured near-dark improves estimates of DC offset, while mid- to high-signal data sharpens conversion-gain estimates--resolving identifiability issues and enabling reliable characterization at any read-noise magnitude. Consequently, the multi-sample extension bridges the traditional PT method and DSERN-specific algorithms \cite{fossum_2015,starkey_2016,gach_2022,nakamoto_2022,hendrickson_2023_PCHEM_theory,Krynski_2025} while delivering parameter-estimate uncertainties comparable to or lower than PT. While this framework improved characterization for emerging DSERN technologies, it nevertheless inherited EM’s susceptibility to local optima.

Ueda and Nakano's Deterministic Annealing (DA) technique offers a principled approach to overcoming local optima in EM by borrowing ideas from statistical physics \cite{ueda_1994_1,ueda_1994_2,ueda_1998}. Rather than optimizing the likelihood directly, DA introduces an annealing parameter, often referred to as ``temperature,'' which deforms the optimization landscape \cite{Zhou_2010}. At high temperature, the objective function is deformed into a form that is simple to optimize. As the temperature is gradually lowered, the original optimization surface is restored, and the algorithm converges to increasingly refined solutions approaching the global optima.

In this work, we extend the PCH-EM framework with quasi-deterministic annealing, yielding a new algorithm we refer to as DA-PCH-EM. This extension retains the analytical tractability of the original PCH-EM algorithm while significantly improving robustness to initialization. We demonstrate through a simulation experiments that DA-PCH-EM provides more stable estimates in the presence of poor initialization, making it well-suited for characterization of next generation of DSERN sensors.

The remainder of the paper is organized as follows. Section \ref{sec:model} provides a brief review of the statistical model driving the PCH-EM algorithm, notation, and the multi-modal nature of likelihood functions generated by DSERN pixels. Section \ref{sec:method} presents the formulation of the DA extension. Section \ref{sec:simulation_results} reports simulation results, and Section \ref{sec:conclusion} concludes with a discussion on implications, limitations, and directions for future work.

%%%%%%%%%%%%%%%%%%%%%%%%%%%%%%%%%%%%%%%%%%%%%%%%%%%%%%%%%%%%%%%%%%%%%%%%%%%%%%%%%%%%
%%%%%%%%%%%%%%%%%%%%%%%%%%%%%%%%%%%%%%%%%%%%%%%%%%%%%%%%%%%%%%%%%%%%%%%%%%%%%%%%%%%%
%%%%%%%%%%%%%%%%%%%%%%%%%%%%%%%%%%%%%%%%%%%%%%%%%%%%%%%%%%%%%%%%%%%%%%%%%%%%%%%%%%%%

\section{Statistical Model and Notation}
\label{sec:model}

Characterization methods for CCD and CMOS image sensors begin with the Photon Counting Distribution (PCD), a Poisson-Gaussian mixture, for describing the stochastic excursions of raw gray counts, a.k.a., Digital Numbers (DN), produced by a pixel under static illumination.  If $X\sim\operatorname{PCD}(H,g,\mu,\sigma)$ denotes the PCD random variable then the associated probability density is:
\begin{equation}
f_X(x|\theta)=\sum_{k=0}^\infty\frac{e^{-H}H^k}{k!}\frac{1}{\sqrt{2\pi}\sigma/g}\exp\left(-\frac{(x-(\mu+k/g))^2}{2(\sigma/g)^2}\right),
\end{equation}
where $H$ is the quanta exposure representing the mean free-electrons generated over the integration time, $g$ is the conversion gain in $(e\text{-}/\mathrm{DN})$, $\mu$ is the DC offset in $(\mathrm{DN})$, $\sigma$ is the input-referred read noise $(e\text{-})$, and $\theta=(H,g,\mu,\sigma)$ is the distribution parameter.

Let $x_{jn}$ represent the $n$th experimentally realized gray count from a pixel with unknown parameter $\theta_j=(H_j,g,\mu,\sigma)$, $\mathbf x_j=(x_{j,1},\dots,x_{j,N_j})$ represent a random sample of size $N_j$, and $\mathbf x=(\mathbf x_1,\dots,\mathbf x_J)$ represent a collection of $J$ random samples captured from the same pixel at varying quanta exposures. Thus, $\mathbf x$ is the experimental data captured for the purpose of characterization. One can then construct the log-likelihood function
\begin{equation}
\label{eq:multi-sample_log-likelihood}
\ell(\boldsymbol\theta|\mathbf x)=\sum_{j=1}^J\sum_{n=1}^{N_j}\log f_X(x_{jn}|\theta_j),
\end{equation}
where $\boldsymbol\theta=(H_1,\dots,H_J,g,\mu,\sigma)$ is the aggregate parameter. Provided an initial guess $\boldsymbol\theta_0$, the PCH-EM algorithm applies a series of closed-form update equations to produce a sequence of parameter estimates $(\boldsymbol\theta_1,\boldsymbol\theta_2,\dots)$ that monotonically increase in log-likelihood until a convergence criteria is met.  The final parameter estimate after convergence is achieved, $\boldsymbol\theta^\ast$, represents the PCH-EM estimate of the unknown aggregate parameter.

\begin{figure}[htb]
    \centering
    \includegraphics[scale=1]{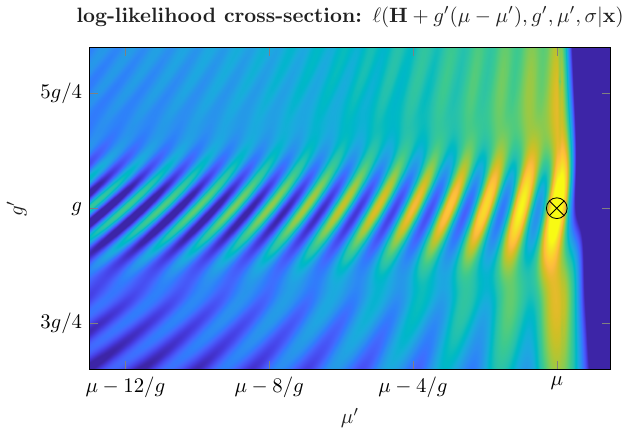}
    \caption{Cross-section of a simulated log-likelihood function showing multiple local maxima. Global maxima marked by a black cross.}
    \label{fig:log_likelihood_cross_section}
\end{figure}

For the DSERN regime ($\sigma<0.5\,e\text{-}$), $\ell$ typically has many local maxima located approximately at $(H_1+m,\dots,H_J+m,g,\mu-m/g,\sigma)$ for $m=0,1,2,\dots$. For example, Figure \ref{fig:log_likelihood_cross_section} plots a cross-section of a two-sample log-likelihood produced from simulated data for the parameter $\boldsymbol\theta=(0.1,3,0.135,200,0.2)$ and sample sizes $\mathbf N=(2000,6000)$. We observe many local maxima with the global maxima, the maxima corresponding to the MLE, marked by a black cross. As the read noise parameter $\sigma$ increases, these local maxima vanish resulting in a unimodal log-likelihood function.

In the context of the PCH-EM algorithm applied to a DSERN pixel, the initial guess $\boldsymbol\theta_0$ must fall within the basin of attraction of the dominant mode of $\ell$, which contains the global maxima.  If the initial guess falls outside of this basin of attraction, the algorithm will converge to a suboptimal mode and return a parameter estimate other than the MLE resulting in an incorrect characterization. When characterizing an array of pixels, it is not possible to evaluate the quality of each solution, thus reducing the initialization dependence of the algorithm can improve reliability and confidence in the solution.

%%%%%%%%%%%%%%%%%%%%%%%%%%%%%%%%%%%%%%%%%%%%%%%%%%%%%%%%%%%%%%%%%%%%%%%%%%%%%%%%%%%%
%%%%%%%%%%%%%%%%%%%%%%%%%%%%%%%%%%%%%%%%%%%%%%%%%%%%%%%%%%%%%%%%%%%%%%%%%%%%%%%%%%%%
%%%%%%%%%%%%%%%%%%%%%%%%%%%%%%%%%%%%%%%%%%%%%%%%%%%%%%%%%%%%%%%%%%%%%%%%%%%%%%%%%%%%

\section{Method: DA-PCH-EM}
\label{sec:method}

Deterministic annealing belongs to the broader class of homotopy-based global optimization methods. The central idea is to introduce a parameterized deformation, or homotopy, that continuously transforms the original multi-modal objective function into a simpler surrogate that is easier to optimize. The optimization is first carried out on this simpler function, and the deformation parameter is then incrementally adjusted to gradually restore the complexity of the original landscape. At each annealing step, the solution obtained from the preceding step serves as the initialization, thereby guiding the search toward increasingly refined solutions. By the final stage, when the homotopy recovers the original objective, the solution typically lies within the basin of attraction of the global optimum, facilitating convergence to the globally optimal solution. The procedure is termed deterministic because the annealing path is prescribed by the deterministic deformation of the objective function rather than stochastic perturbations. For DA-PCH-EM, we take a quasi-deterministic variant of this approach to improve the robustness of the PCH-EM algorithm in the presence of poor initialization and increase the probability of obtaining MLEs of the pixel performance parameters.

To construct the homotopy function for DA-PCH-EM note that in the DSERN regime, the density of $X\sim\operatorname{PCD}(H,g,\mu,\sigma)$ exhibits many distinct Gaussian peaks (local maxima) with spacing $1/g$. The multi-modality of the log-likelihood function is a direct consequence of the many local maxima in the density of $X$ at low read noise. Since $X$ is a location-scale family for the standard Poisson-Gaussian mixture distribution, adding Gaussian noise does not change the mathematical formulation of the distribution. Thus we define
\begin{equation}
\label{eq:annealed_PCD_variable}
X_\beta=X+a(\beta)Z,
\end{equation}
where $\beta\in[0,1]$ is the ``temperature'', $a(\cdot)$ is a nonnegative function with the properties $a(0)=0$ and $\lim_{\beta\to 1^-}=\infty$, $Z\sim\operatorname{Gaussian}(0,1)$ is a standard normal variable independent of $X$. Consequently, $X_\beta\sim\operatorname{PCD}(H,g,\mu,\sigma_\mathrm{eff}(\beta))$, where $\sigma_\mathrm{eff}^2(\beta)=\sigma^2+g^2a^2(\beta)$ is the temperature dependent effective read noise. In this form we observe that $X_0$ (minimum temperature) recovers the original PCD variable and as $\beta\to 1^-$ (high temperature), $X_\beta\overset{d}{\to}\operatorname{Gaussian}(\mu+H/g,(H+\sigma^2)/\sigma_\mathrm{eff}(\beta)^2)$ resulting in a unimodal Gaussian distribution. This last property can be deduced by showing the limit applied to the characteristic function of $(X_\beta-\mathsf EX_\beta)/\sqrt{\mathsf{Var}X_\beta}$ yields the characteristic function of the standard Gaussian distribution.  Thus, supplementing $X$ with temperature controlled Gaussian noise accomplishes two things: 1) keeps $X_\beta$ in the PCD family and 2) removes the multimodal structure in the density of $X$ (and $\ell$) at high temperatures.  This leads us to construct the homotopy
\begin{equation}
\mathcal H(\boldsymbol\theta,\beta|\mathbf x,\mathbf z)=\ell(\boldsymbol\theta|\mathbf x+a(\beta)\mathbf z),
\end{equation}
where $\ell(\cdot)$ is the log-likelihood function given in (\ref{eq:multi-sample_log-likelihood}) and $\mathbf z$ is a pseudo-random sample of standard normal variates the same size as the experimental pixel data $\mathbf x$.

Upon inspection, the constructed homotopy $\mathcal H$ is mathematically equivalent to the original log-likelihood function in (\ref{eq:multi-sample_log-likelihood}) with effective read noise of $\sigma_\mathrm{eff}$; therefore, the PCH-EM algorithm can optimize it for any $\beta$ without modification. Algorithm \ref{alg:dapchem} presents the DA-PCH-EM algorithm. The algorithm requires the experimental data $\mathbf x$ and initial guess $\boldsymbol\theta_0$ just as the standard algorithm does with the addition of the stochastic perturbation data $\mathbf z$.  The inclusion of perturbation data is why we classify the algorithm as quasi-deterministic since the annealing path is deterministic once the stochastic perturbation data is realized. In addition, DA-PCH-EM requires an annealing function $a(\cdot)$, maximum temperature $\beta_{\max}$, and number of annealing steps $M$. These three additional requirements together determine the annealing schedule for the algorithm. As previously described, the algorithm runs the standard PCH-EM algorithm on the perturbed data $\mathbf x_m$ for a sequence of decreasing temperatures and then returns the final parameter estimate $\boldsymbol\theta_\mathrm{da}^\ast$.
\begin{algorithm}[htb]
\caption{\texttt{DA-PCH-EM}}
\label{alg:dapchem}
\begin{algorithmic}[1]
\REQUIRE $\mathbf x$ (experimental data), $\mathbf z$ (synthetic Gaussian perturbations), $\boldsymbol\theta_0$ (initial guess), $a(\cdot)$ (annealing function), $\beta_{\max}$ (max.~temperature), $M$ (number of annealing steps)
\STATE Generate sequence of $M$ uniformly spaced and decreasing $\beta$-values from $\beta_{\max}$ to zero: $(\beta_1,\dots,\beta_M)=(\beta_{\max},\dots,0)$.
\FOR{$m=1$ to $M$}
\STATE Perturb data: $\mathbf x_m=\mathbf x+a(\beta_m)\mathbf z$.
\STATE Update solution: Obtain $\boldsymbol\theta_m$ by running PCH-EM algorithm on $\mathbf x_m$ for initial guess $\boldsymbol\theta_{m-1}$.
\ENDFOR
\STATE $\boldsymbol\theta_\mathrm{da}^\ast=\boldsymbol\theta_M$
\RETURN $\boldsymbol\theta_\mathrm{da}^\ast$
\end{algorithmic}
\end{algorithm}

%%%%%%%%%%%%%%%%%%%%%%%%%%%%%%%%%%%%%%%%%%%%%%%%%%%%%%%%%%%%%%%%%%%%%%%%%%%%%%%%%%%%
%%%%%%%%%%%%%%%%%%%%%%%%%%%%%%%%%%%%%%%%%%%%%%%%%%%%%%%%%%%%%%%%%%%%%%%%%%%%%%%%%%%%
%%%%%%%%%%%%%%%%%%%%%%%%%%%%%%%%%%%%%%%%%%%%%%%%%%%%%%%%%%%%%%%%%%%%%%%%%%%%%%%%%%%%

\section{Results \& Analysis}
\label{sec:simulation_results}

We now demonstrate the effectiveness of the proposed DA-PCH-EM algorithm through a simulated example using the simulation parameters given in Table \ref{tbl:sim_parameters}. The experimental and pixel parameters are the same used to create Figure \ref{fig:log_likelihood_cross_section} to ensure the log-likelihood function is multimodal.  The pixel parameters were chosen to be similar to published values for the DSERN capable Hamamatsu Photonics Orca Quest QIS. The logarithmic annealing function was chosen to match the required properties for the DA-PCH-EM algorithm but also to roughly approximate the time dependent cooling of a hot object to ambient temperature (rapid cooling at high temperature followed by gradual cooling at lower temperature).  This choice not only serves to approximate natural cooling but also is computationally efficient as the PCH-EM algorithm is most computationally expensive, requiring many iterations to achieve convergence, at high temperatures \cite{Naim_2012}. The the maximum temperature, $\beta_{\max}$, was chosen to ensure the homotopy function is sufficiently ``flat'' at the maximum temperature to avoid multiple local optima in the early annealing steps.  To see how this determination is made, recall (\ref{eq:annealed_PCD_variable}) where the effective read noise as a function of temperature can be written as
\begin{equation}
\sigma_\mathrm{eff}(\beta)=(\sigma^2+g^2a^2(\beta))^{1/2}.
\end{equation}
To ensure the homotopy $\mathcal H$ lacks local maxima, one wants to select $\beta_{\max}$ such that $\sigma_\mathrm{eff}(\beta_{\max})$ exceeds the DSERN threshold, i.e., $\sigma_\mathrm{eff}(\beta_{\max})\gg 0.5\,e\text{-}$. In this example, $\sigma_\mathrm{eff}(\beta_{\max})=0.81$, which satisfies the requirement.  In practice, a rough estimate of $g$ determined from peak spacing in the histogram of a single image frame and a lower bound on $\sigma$ can be combined to produce a single maximum temperature able characterize an array of pixels with varying read noise magnitudes.
\begin{table}[htb]
\caption{Simulation parameters.}
\label{tbl:sim_parameters}
\centering
\begin{tabular}{ |p{4cm}|p{2.5cm}|p{2.5cm}|  }
\hline
\textbf{Parameter}& \textbf{Value} &\textbf{Unit} \\
\hline
\multicolumn{3}{|c|}{Experimental Parameters} \\
\hline
$\mathbf N$ (sample sizes)& $(2000,6000)$ &$-$\\
$\mathbf H$ (quanta exposures)& $(0.1,3)$ &$e\text{-}$\\
\hline
\multicolumn{3}{|c|}{Pixel Parameters} \\
\hline
$g$ (conversion gain)& $0.135$ &$e\text{-}/\mathrm{DN}$\\
$\mu$ (DC offset)& $200$ &$\mathrm{DN}$\\
$\sigma$ (read noise)& $0.2$ &$e\text{-}$ \\
\hline
\multicolumn{3}{|c|}{Annealing Schedule} \\
\hline
$a(\beta)$ (annealing function)& $-\log(1-\beta)$ &$-$\\
$\beta_{\max}$ (max.~temperature)& $0.997$ &$-$\\
$M$ (\# of annealing steps)& $10$ &$-$ \\
\hline
\hline
\end{tabular}
\end{table}

Procedure \ref{proc:daPCH-EM_simulation} provides the simulation steps that were implemented in the MATLAB programming environment.  It begins by generating a synthetic experimental dataset $\mathbf x$ in accordance with the PCD model and parameters in Table \ref{tbl:sim_parameters}, along with the standard normal perturbations $\mathbf z$.  The procedure then repeats the process of generating a randomized initial guess, computing the PCH-EM solution, and then updating the PCH-EM solution using the DA extension a total of $N_\mathrm{trials}$ trials. By seeding DA-PCH-EM with the PCH-EM solution, the algorithm is stress tested since it's initialized at a local optima of the log-likelihood function. For each trial, the log-likelihood of both solutions is evaluated to determine if the annealing was able to improve the basline solution.

To simulate the randomized initialization, five standard uniform pseudo-random variates $\mathbf U=(U_1,\dots,U_5)$, $U_i\overset{\mathrm{iid}}{\sim}\operatorname{Uniform}(0,1)$, are generated and the initial guess $\boldsymbol\theta_0=(H_{1,0},H_{2,0},g_0,\mu_0,\sigma_0)$ is computed according to
\begin{equation}
\label{eq:random_initial_guess}
\begin{aligned}
H_{j,0} &=\max(0.05,H_j(U_j+0.5))\\
g_0 &=g(U_3+0.5)\\
\mu_0 &=\mu-(10-12\,U_4)/g\\
\sigma_0 &=\sigma(U_5+0.5).
\end{aligned}
\end{equation}
This rule for generating starting points was chosen to intentionally initialize the algorithm for parameters near local optima of the likelihood function. In this manner, the simulation can evaluate the robustness of both algorithms for a large span of initial guesses on a single dataset $\mathbf x$.

\begin{procedure}[htb]
\caption{Compare \texttt{PCH-EM} to \texttt{DA-PCH-EM}}
\label{proc:daPCH-EM_simulation}
\begin{algorithmic}[1]
\STATE Generate $\mathbf x$ and $\mathbf z$.
\FOR{$n=1$ to $N_\mathrm{trials}$}
\STATE Random initialization: Generate $\mathbf U$ and compute $\boldsymbol\theta_0$ according to (\ref{eq:random_initial_guess}).
\STATE PCH-EM Solution: Compute the PCH-EM solution, $\boldsymbol\theta^\ast$, using the initial guess $\boldsymbol\theta_0$.
\STATE DA-PCH-EM Update: Compute the DA-PCH-EM solution, $\boldsymbol\theta^\ast_\mathrm{da}$, using the initial guess $\boldsymbol\theta^\ast$.
\STATE Evaluate log-likelihood of each solution: $\mathbf L_n=(\ell(\boldsymbol\theta^\ast|\mathbf x),\ell(\boldsymbol\theta^\ast_\mathrm{da}|\mathbf x))$.
\ENDFOR
\RETURN $\mathbf L_1,\dots,\mathbf L_{N_\mathrm{trials}}$
\end{algorithmic}
\end{procedure}

%%%%%%%%%%%%%%%%%%%%%%%%%%%%%%%%%%%%%%%%%%%%%%%%%%%%%%%%%%%%%%%%%%%%%%%%%%%%%%%%%%%%

\subsection{Simulation Results}

\begin{figure}[htb]
    \centering
    \includegraphics[scale=1]{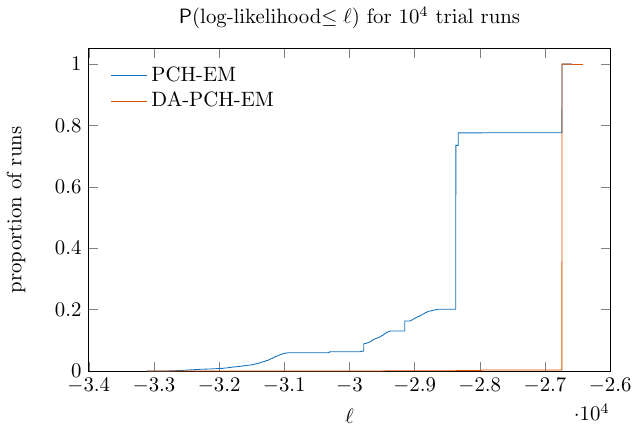}
    \caption{Cumulative distribution of log-likelihood achieved by $10^4$ PCH-EM and DA-PCH-EM solutions.}
    \label{fig:simulation_results}
\end{figure}

The simulation outlined in Procedure \ref{proc:daPCH-EM_simulation} was executed for $N_\mathrm{trials}=10^4$ trials on a Lenovo Legion Pro 5 laptop equipped with an Intel Core i9-14900HX processor. The simulation took a total of $457.3$ seconds to complete resulting in a net cost of $0.046$ seconds per trial. Figure \ref{fig:simulation_results} plots the cumulative distribution of log-likelihood obtained by the PCH-EM and DA-PCH-EM solutions.  Discontinuous jumps in the cumulative distribution of the PCH-EM data correspond to large numbers of trials where PCH-EM became trapped in highly attractive but nondominant modes of the log-likelihood function. The maximum log-likelihood for all trials across both methods was $\ell=-26738.9$. Of the $10^4$ trials performed, $77.55\%$ of the PCH-EM solutions failed to achieve this maximum value compared to only $0.27\%$ for the DA-PCH-EM solutions. This demonstrates not only the susceptibility of PCH-EM to becoming trapped in local optima but also the dramatic improvement in robustness to poor initialization achieved with the deterministic annealing extension.

%%%%%%%%%%%%%%%%%%%%%%%%%%%%%%%%%%%%%%%%%%%%%%%%%%%%%%%%%%%%%%%%%%%%%%%%%%%%%%%%%%%%
%%%%%%%%%%%%%%%%%%%%%%%%%%%%%%%%%%%%%%%%%%%%%%%%%%%%%%%%%%%%%%%%%%%%%%%%%%%%%%%%%%%%
%%%%%%%%%%%%%%%%%%%%%%%%%%%%%%%%%%%%%%%%%%%%%%%%%%%%%%%%%%%%%%%%%%%%%%%%%%%%%%%%%%%%

\section{Conclusion}
\label{sec:conclusion}

This work introduced DA-PCH-EM, a quasi-deterministic annealing extension of the PCH-EM algorithm that retains the tractability of the original framework while improving robustness to initialization.  A simulation was performed on synthetic data demonstrating a significant improvement in the robustness of the proposed DA-PCH-EM algorithm to poor initialization over the non-annealing variant. The ability to reliably estimate the intrinsic performance parameters of imaging pixels is not only important for device characterization but, for DSERN capable devices, for accurate calibration of arbitrary raw gray counts to physical counts of electrons. As of now, the DA extension accomplishes a number of important feats by providing a robust characterization method capable of characterizing pixels at any read noise magnitude with as good or better performance over the legacy Photon Transfer method.  However, a key limitation common to all current characterization methods still exists: the linearity of the underlying statistical model.  In CMOS pixels, the signal dependent capacitance of the sense node results in a signal dependent, non-constant, conversion gain.  Given that the statistical model driving the DA-PCH-EM algorithm assumes a constant conversion gain, this limits the algorithm to characterizing the pixel over only a subset of its dynamic range where conversion gain can be approximated as constant. As such, future work should focus on generalizing the model to incorporate signal dependent conversion gain. By doing so, DSERN capable devices can be characterized over their full dynamic range resulting in accurate quanta counting for a large span of signal levels that are critical to photon number resolving applications.

%%%%%%%%%%%%%%%%%%%%%%%%%%%%%%%%%%%%%%%%%%%%%%%%%%%%%%%%%%%%%%%%%%%%%%%%%%%%%%%%%%%%
%%%%%%%%%%%%%%%%%%%%%%%%%%%%%%%%%%%%%%%%%%%%%%%%%%%%%%%%%%%%%%%%%%%%%%%%%%%%%%%%%%%%
%%%%%%%%%%%%%%%%%%%%%%%%%%%%%%%%%%%%%%%%%%%%%%%%%%%%%%%%%%%%%%%%%%%%%%%%%%%%%%%%%%%%

\section*{Acknowledgment}
Aaron Hendrickson thanks the Naval Innovative Science \& Engineering (NISE) executive board for funding support under project no. 219BAR-24-043.

\bibliographystyle{unsrt}
\bibliography{sources}

\end{document}